\documentclass[aps,prl,preprint, showpacs, superscriptaddress]{revtex4}

\usepackage[dvips]{graphicx}
\usepackage{latexsym}

\begin{document}

  \title{Numerical evaluation of the upper critical dimension of percolation \\
    in scale-free networks}
  
  \author{Zhenhua Wu}
  \affiliation{Center for Polymer Studies, Boston University, Boston,
    Massachusetts 02215, USA}
  \author{Cecilia Lagorio}
  \affiliation{Departamento de F\'{\i}sica, Facultad de Ciencias Exactas y
    Naturales, Universidad Nacional de Mar del Plata, Funes 3350, 7600 Mar del
    Plata, Argentina}
  \author{Lidia A. Braunstein}
  \affiliation{Center for Polymer Studies, Boston University, Boston,
    Massachusetts 02215, USA}
  \affiliation{Departamento de F\'{\i}sica, Facultad de Ciencias Exactas y
    Naturales, Universidad Nacional de Mar del Plata, Funes 3350, 7600 Mar del
    Plata, Argentina}
  \author{Reuven Cohen}
  \author{Shlomo Havlin}
  \affiliation{Minerva Center of Department of Physics, Bar-Ilan
    University, Ramat Gan, Israel}
  \author{H. Eugene Stanley}
  \affiliation{Center for Polymer Studies, Boston University, Boston,
    Massachusetts 02215, USA}
  
  \date{\today}
  
  \begin{abstract}
    
    We propose a numerical method to evaluate the upper critical dimension
    $d_c$ of random percolation clusters in Erd\H{o}s-R\'{e}nyi networks and
    in scale-free networks with degree distribution ${\cal P}(k) \sim
    k^{-\lambda}$, where $k$ is the degree of a node and $\lambda$ is the
    broadness of the degree distribution. Our results report the theoretical
    prediction, $d_c = 2(\lambda - 1)/(\lambda - 3)$ for scale-free networks
    with $3 < \lambda < 4$ and $d_c = 6$ for Erd\H{o}s-R\'{e}nyi networks and
    scale-free networks with $\lambda > 4$. When the removal of nodes is not
    random but targeted on removing the highest degree nodes we obtain $d_c =
    6$ for all $\lambda > 2$. Our method also yields a better numerical
    evaluation of the critical percolation threshold, $p_c$, for scale-free
    networks. Our results suggest that the finite size effects increases when
    $\lambda$ approaches $3$ from above..
  \end{abstract}
  

\maketitle

Recently much attention has been focused on the topic of complex networks,
which characterize many natural and man-made systems, such as the Internet,
airline transport system, power grid infrastructures, and the world wide web
(WWW)~\cite{Barabasi_rmp_review, vespignani_book, Mendes_book,
Cohen_book}. Many studies on these systems reveal a common power law degree
distribution, ${\cal P}(k) \sim k^{-\lambda}$ with $k \ge k_{\rm min}$, where
$k$ is the degree of a node, $\lambda$ is the exponent quantifying the
broadness of the degree distribution~\cite{Barabasi_science_99} and $k_{\rm
min}$ is the minimum degree. Networks with power law degree distribution are
called scale-free (SF) networks. The power law degree distribution represents
topological heterogeneity of the degree in SF networks resulting in the
existence of hubs that connect significant fraction of nodes. In this sense,
the well studied Erd\H{o}s-R\'{e}nyi (ER) networks~\cite{erdos, erdos_2,
Bollobas} are homogeneous and can be represented by a characteristic degree
$\langle k \rangle$, the average degree of a node, while SF networks are
heterogeneous and do not have a characteristic degree.

The embedded dimension of ER and SF networks can be regarded as infinite ($d
= \infty$) since the number of nodes within a given ``distance'' increases
exponentially with the distance compared to an Euclidean $d$ dimensional
lattice network where the number of nodes within a distance $L$ scales as
$L^d$. Percolation theory is a powerful tool to describe a large number of
systems in nature such as porous and amorphous materials, random resistor
networks, polymerization process and epidemic spreading and immunization in
networks~\cite{Bunde_book, stauffer_book}. Percolation theory study the
topology of a network of $N$ nodes resulting from removal of a fraction $q
\equiv 1 - p$ of nodes (or links) from the system. It is found that in general
there exists a critical phase transition at $p = p_c$, where $p_c$ is the
critical percolation threshold. Above $p_c$, most of the nodes (order $N$)
are connected, while below $p_c$ the network collapses into small clusters of
sizes of order $\ln N$. For lattices in $d \ge 6$, the nodes, in the
percolation cluster, do not have spatial constraints and therefore all
percolation exponents remain the same and the system behavior can be
described by mean field theory~\cite{stauffer_book, Bunde_book}. This is
because at $d_c = 6$ the spatial constraints on the percolation clusters
become irrelevant and each shortest path between two nodes in the percolation
cluster at criticality can be considered as a random walk. The critical
dimension $d_c$ above which the critical exponents of percolation become the
same as in mean field theory is called the {\it upper critical dimension}
(UCD). It is well known that the UCD for percolation in d-dimensional
lattices is $6$.  Studies of percolation in ER networks, yield the same
critical exponents as in mean-field values of regular percolation in infinite
dimensions. This is because in ER networks spatial constraints do not appear
and the symmetry is almost the same as in Euclidean lattices, i.e., there is
a typical number of links per node. However, SF networks with $2 < \lambda <
4$ have different critical exponents than ER networks~\cite{Cohen_exponents,
Cohen_chapter}. The regular mean-field exponents are recovered only for SF
networks with $\lambda > 4$. This is due to the fact that for the classical
mean field one needs two conditions (a) no spatial constraint (b)
translational symmetry, meaning that all nodes have similar neighborhood. The
second condition does not apply for SF networks with $\lambda < 4$ due to the
broad degree distribution and thus we expect a new type of mean field
exponents~\cite{Cohen_book}. Indeed, for SF networks with $3 < \lambda < 4$,
the UCD was shown to be~\cite{Cohen_chapter}:
\begin{equation}
  d_c \equiv \frac{2(\lambda -1)}{\lambda - 3}.
  \label{ucd_sf}
\end{equation}
Thus, $d_c$ is larger than $6$ and for $\lambda \to 3$, $d_c \to \infty$.
When scale-free networks are embedded in a regular Euclidean
lattice~\cite{Manna, Rozenfeld, Waren}, the value of $d_c$ tells us above
which dimension the percolation clusters will not be affected by the spatial
constraints and therefore the percolation exponents will be the same as for
infinite dimension. Thus, it is reasonable that when $\lambda$ is smaller,
the network is more complex (due to bigger hubs) and a higher upper critical
dimension is expected. However, Eq.~(\ref{ucd_sf}), that was shown
analytically to be valid for $N \to \infty$ was never verified or tested
numerically. It is also interesting to determine the range of $N$ values
where the results of Eq.~(\ref{ucd_sf}) can be observed. Here we propose a
numerical method to measure the value of $d_c$ for ER and SF networks with
$\lambda > 3$~\cite{FN_1}.

Finite-size scaling arguments in $d$-dimensional lattice networks
predict~\cite{Bunde_book, stauffer_book} that the critical threshold $p_c(L)$
approaches $p_c \equiv p_c(\infty)$ via,
\begin{equation}
  p_c(L) - p_c(\infty) \sim L^{-1/\nu}, 
  \label{finite_size_scale_lattice}
\end{equation}
where $L$ is the linear lattice size and $\nu$ is the correlation critical
exponent.
Eq.~(\ref{finite_size_scale_lattice}) for lattices can be generalized to
networks of $N$ nodes via the relation $L^d = N$, i.e., $p_c(N) - p_c(\infty)
\sim N^{(-1 / d \nu)}$. Since networks can be regarded as embedded in infinite
dimension and since above $d_c$ all exponents are the same, we replace $d$ by
$d_c$,
\begin{equation}
  p_c(N) - p_c(\infty) \sim N^{-1/ d_c \nu} \equiv N^{-\Theta}.
\label{finite_size_scale}
\end{equation}
 For ER and SF networks with $\lambda > 4$, we have $d_c = 6$ and $\nu =
1/2$, thus from Eq.~(\ref{finite_size_scale}) follows,
\begin{equation}
  p_c(N) - p_c(\infty) \sim N^{- 1/3}.
\label{finite_size_scale_er}
\end{equation}
For SF networks with $3 < \lambda < 4$, we have $\nu = 1/2$ and substituting
Eq.~(\ref{ucd_sf}) in Eq.~(\ref{finite_size_scale}), it yield,
\begin{equation}
  p_c(N) - p_c(\infty) \sim N^{(3 - \lambda)/(\lambda - 1)}.
\label{finite_size_scale_detail}
\end{equation}
In this paper we use Eq.~(\ref{finite_size_scale}) to measure $\Theta \equiv
2 / d_c$ from which we can evaluate $d_c$. To measure $\Theta$, using the
finite size scaling of Eq.~(\ref{finite_size_scale}), we have to compute the
dependence of the percolation threshold, $p_c(N)$, of ER and SF networks on
the system size $N$. To calculate $p_c(N)$, we apply the second largest
cluster method~\cite{Bunde_book, stauffer_book}, which is based on
determining $p_c(N)$ by measuring the value of $p_c$ at the maximum value of
the average size of the second largest cluster, $\langle S_{2} \rangle$. It
is known that $\langle S_{2} \rangle$ has a sharp peak as a function of
$p$ at $p_c$~\cite{Bunde_book, stauffer_book}. To detect this peak we perform
a Gaussian fit around the peak and estimate the peak position which is
$p_c(N)$~\cite{FN_kappa}.

%
To improve the speed of the simulations, we implement the fast Monte Carlo
algorithm for percolation proposed by Newman and
Ziff~\cite{Newman_fast_algo}. Basically, for each realization, we prepare one
instance of $N$ nodes network with the desired structure as the reference
network. Then we prepare another set of $N$ nodes with no links as our target
network. Because we want to know the size of the 2nd largest cluster instead
of the largest one, we use a list which keeps track of all the clusters in
descending order according to their sizes, which in the beginning is a list
of $N$ clusters of size one. As we choose the links in random order from the
reference network and make the connection in the target network, we update
the list of the cluster size but always keep them in descending order. The
concentration value, $p$, of each newly connected link is calculated by the
number of links after adding this link in the target network divided by the
total number of links in the reference network. We record $S_2$ in the
following way. First, we make $1000$ bins between $0$ and $1$. When each link
is connected, we record $S_2$ at the concentration value $p$ of this
newly connected link. After many realizations, we take the average of $S_2$
for each bin.

\begin{figure}
  \includegraphics[width=0.19\textwidth, angle=-90]{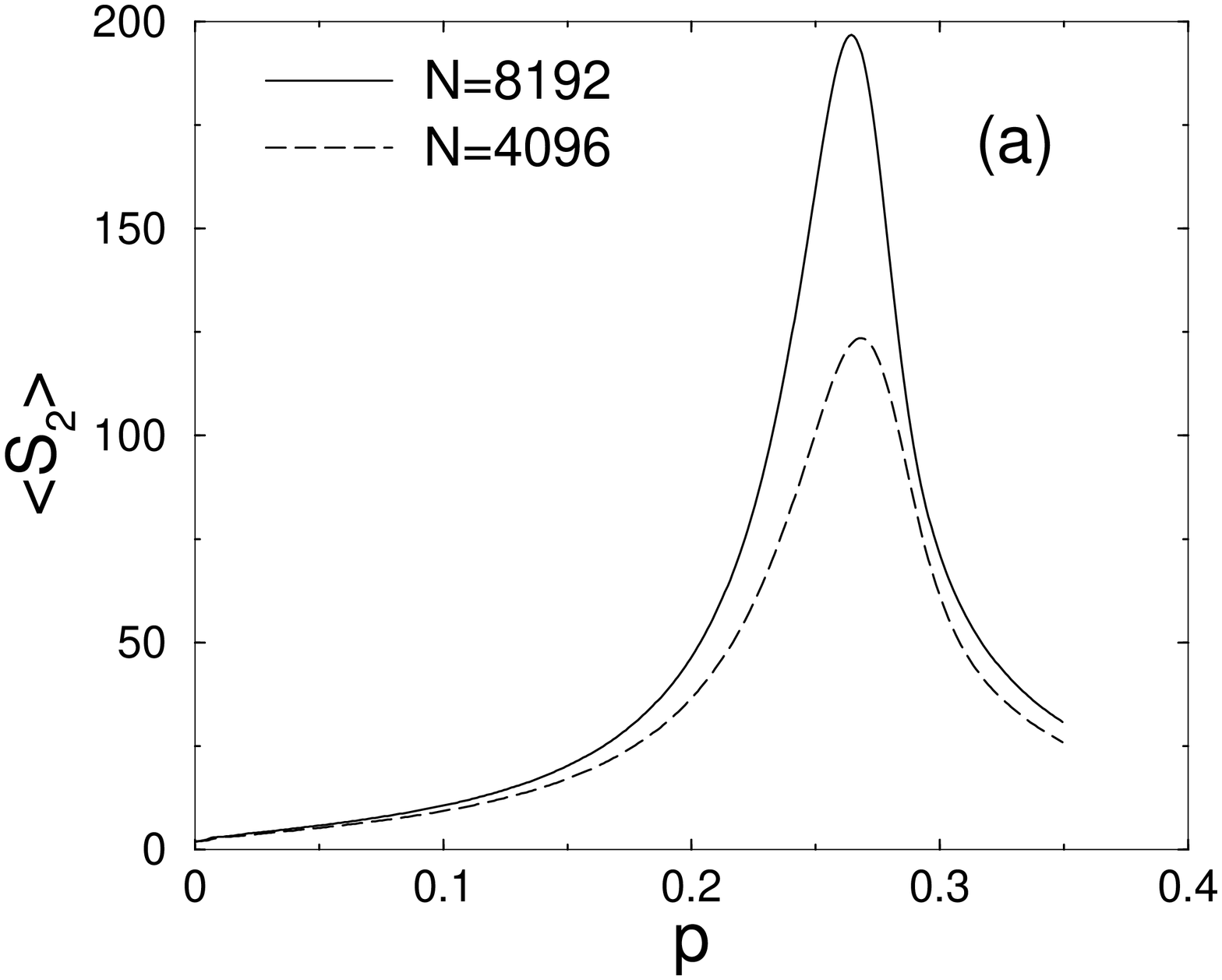}
  \includegraphics[width=0.19\textwidth, angle=-90]{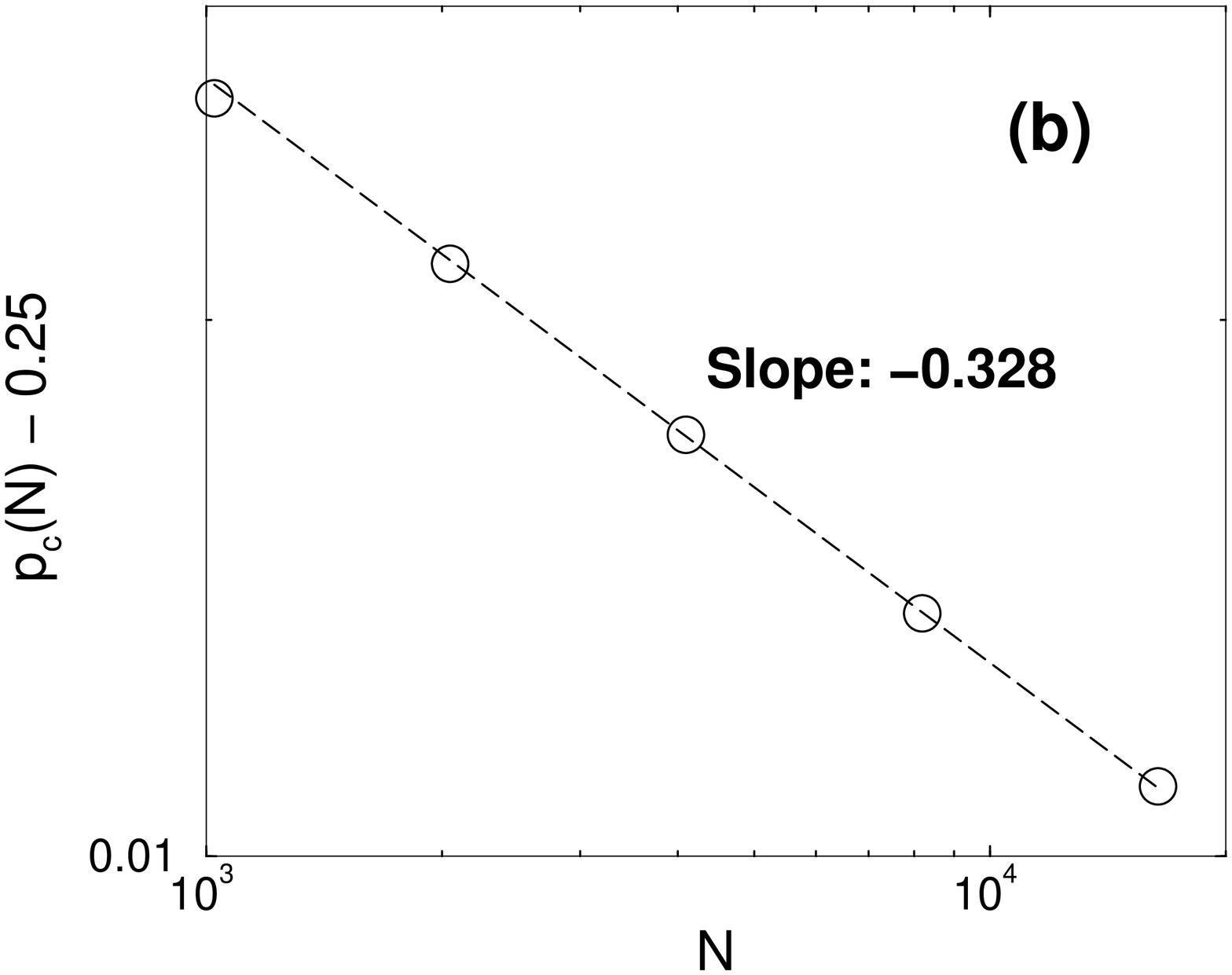}
  \caption{(a) The average size of the 2nd largest cluster, $\langle S_2
    \rangle$, as a function of the concentration, $p$, of links present in
    the ER networks. The typical number of realizations for each curve is
    $10^6$. (b) Log-log plot of $p_c(N) - p_c(\infty)$ as a function of $N$,
    where $p_c(\infty) = 1 / \langle k \rangle = 0.25$ for ER with $\langle k
    \rangle = 4$.}
  \label{s2_pcN_er.graph}
\end{figure}

Figure~\ref{s2_pcN_er.graph}(a) shows $\langle S_2 \rangle$ as a function of
$p$, for two different system sizes of ER networks with $\langle k \rangle =
4$. The position of the peak, obtained by fitting the peak with a Gaussian
function, yields $p_c(N)$. Figure~\ref{s2_pcN_er.graph}(b) shows $p_c(N)$ as
a function $N$. Using $p_c(\infty) \equiv 1/\langle k \rangle$ =
0.25~\cite{erdos, erdos_2}, the fitting of Eq.~(\ref{finite_size_scale})
gives the exponent $\Theta = 0.328 \pm 0.003$, very close to the theoretical
prediction for ER, $\Theta = 1/3$, Eq.~(\ref{finite_size_scale_er}). We
performed the same simulations for ER with other average degrees, $\langle k
\rangle = 5$ and $6$, and obtained similar results for $\Theta$.

To determine $p_c(\infty)$ for random SF networks, we use the exact
analytical results~\cite{Cohen_robust},
\begin{equation}
  p_c(\infty) \equiv \frac{1}{\kappa_0 - 1}.
  \label{pc_sf_cohen}
\end{equation}
Here $\kappa_0 \equiv \langle k_0^2 \rangle / \langle k_0 \rangle$ is
computed from the original degree distribution (${\cal P}(k_0)$) for which
the network is constructed. However, the way to compute the value of
$\kappa_0$ is strongly affected by the algorithm of generating the SF network
as explained below.

To generate SF networks with power law exponent $\lambda$, we use the
Molloy-Reed algorithm~\cite{Molloy_Reed_book, Molloy_Reed}. We first generate
a series of random real number $u$ satisfying the distribution ${\cal P}(u) =
c u^{-\lambda}$, where $c = (\lambda - 1)/k_{\rm min}^{1 - \lambda}$ is the
normalization factor. Next we truncate the real number $u$ to be an integer
number $k$, which we assume to be the degree of a node. We make $k$ copies of
each node according to its degree and randomly choose two nodes and connect
them by a link. Notice that the process of truncating the real number $u$ to
be an integer number $k$ which is the degree of a node actually slightly
changes the degree distribution because any real number $n \leq u < n + 1$,
where $n$ is an integer number, will be truncated to be equal $n$. Thus, the
actual degree distribution we obtain using this algorithm is
\begin{equation}
  {\cal P}(k) = \int_{k}^{k+1}c u^{-\lambda}du = \frac{1}{k_{\rm min}^{1 -
      \lambda}}(k^{1 - \lambda} - (k + 1)^{1 - \lambda}).
  \label{pk_modified}
\end{equation}
We use Eq.~(\ref{pk_modified}) to compute $\kappa_0$ and $p_c(\infty)$
defined in Eq.~(\ref{pc_sf_cohen}). Table~\ref{table_results} shows the
calculated results of $p_c(\infty)$ for several values of $\lambda$.

\begin{figure}
  \includegraphics[width=0.19\textwidth, angle=-90]{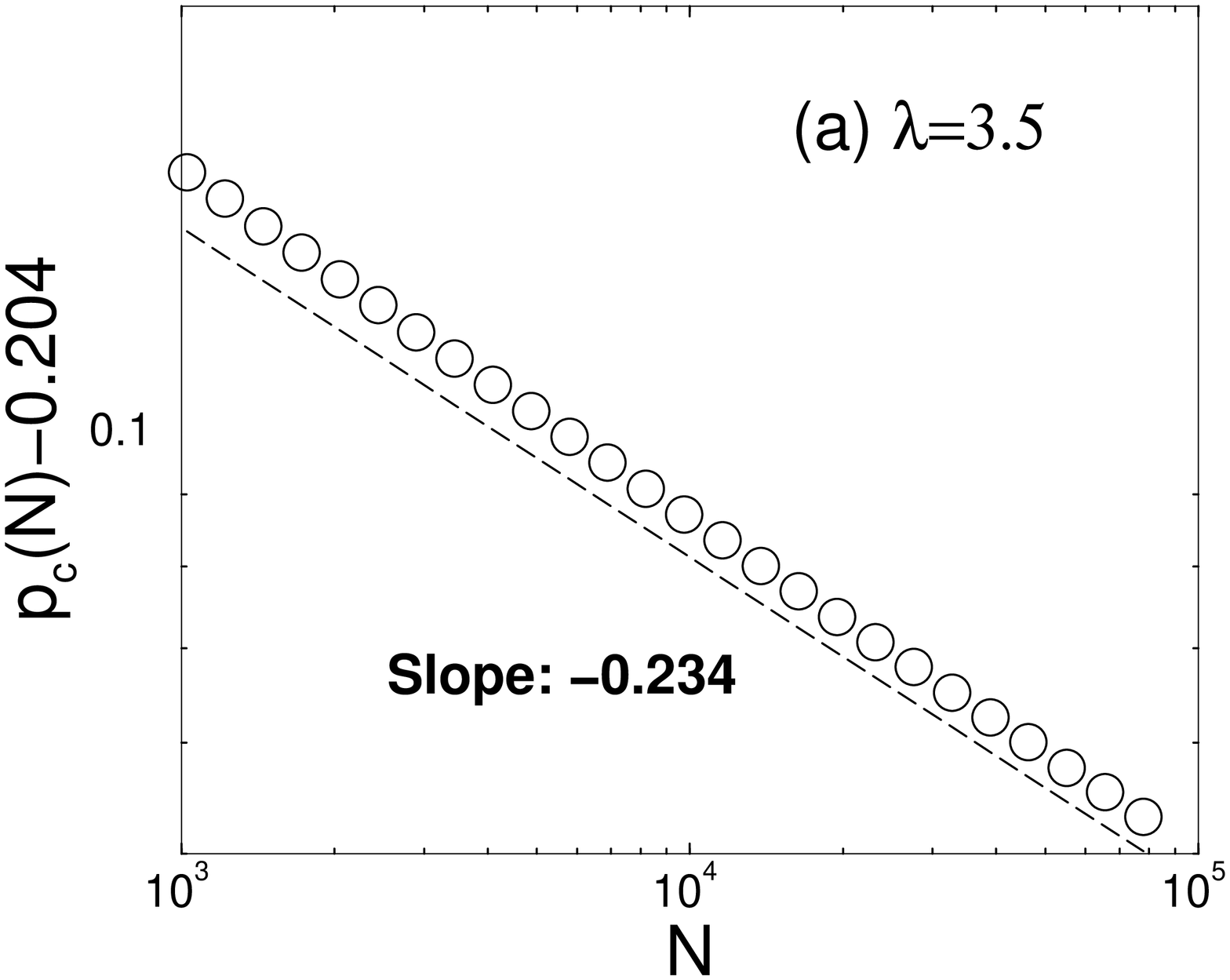}
  \includegraphics[width=0.19\textwidth, angle=-90]{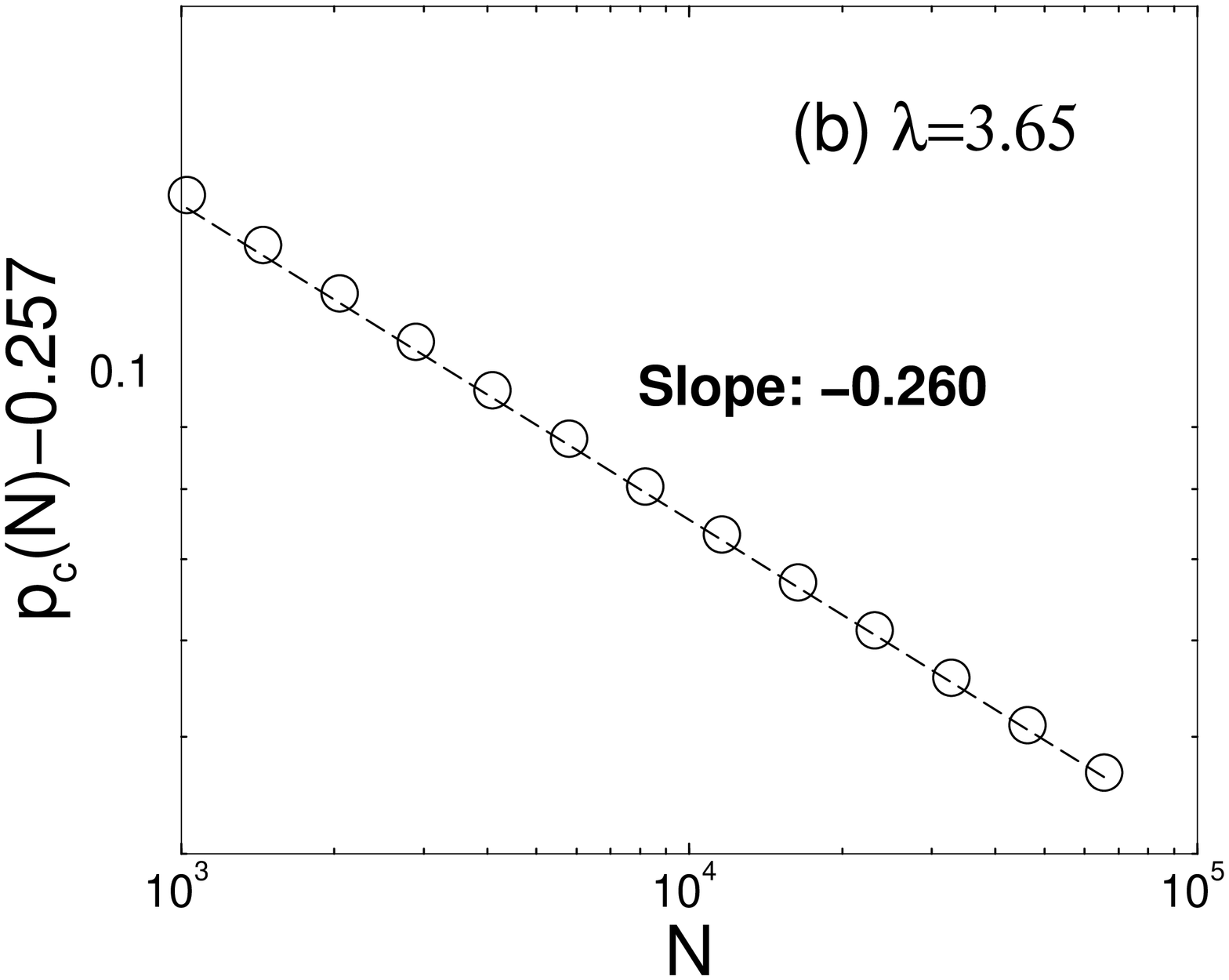}
  \includegraphics[width=0.19\textwidth, angle=-90]{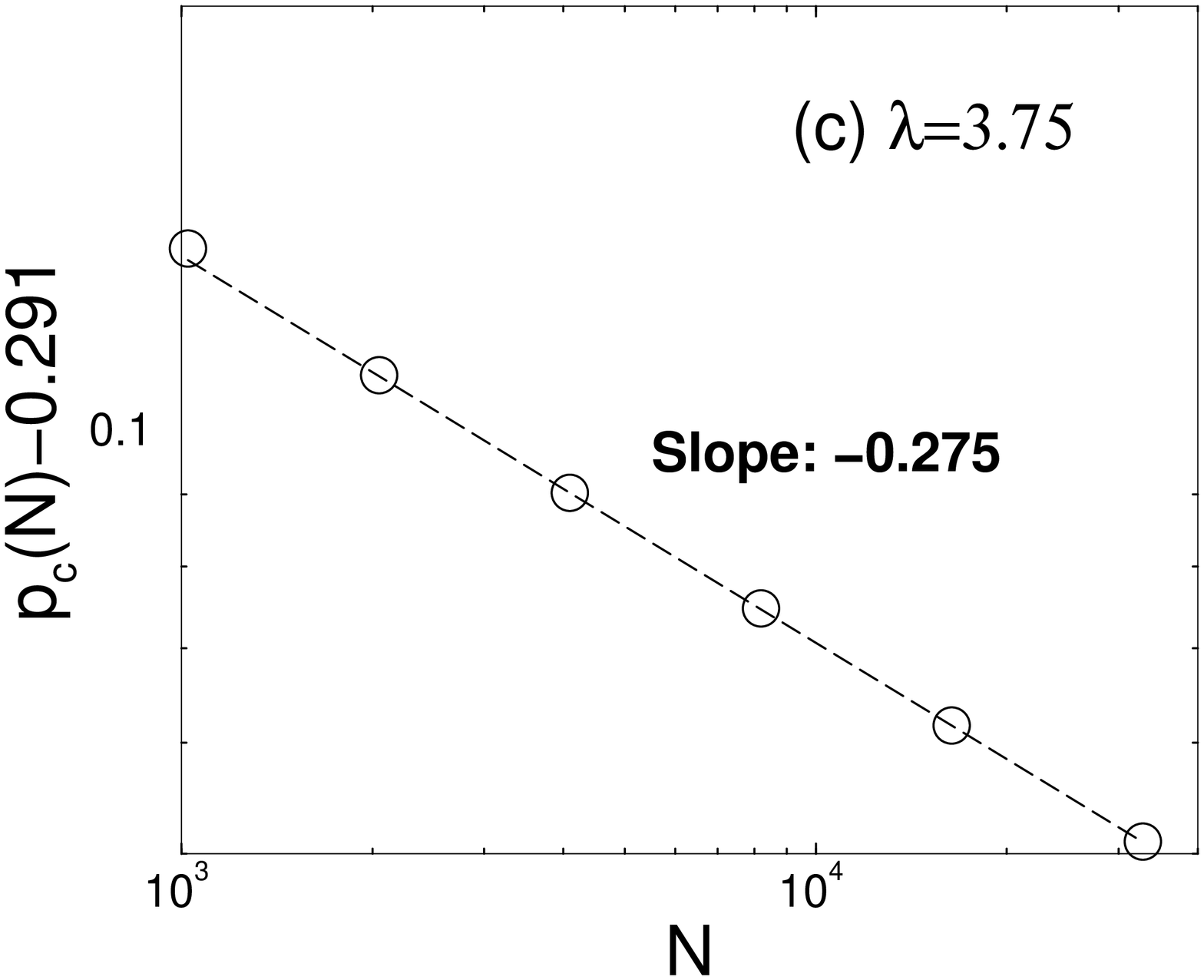}
  \includegraphics[width=0.19\textwidth, angle=-90]{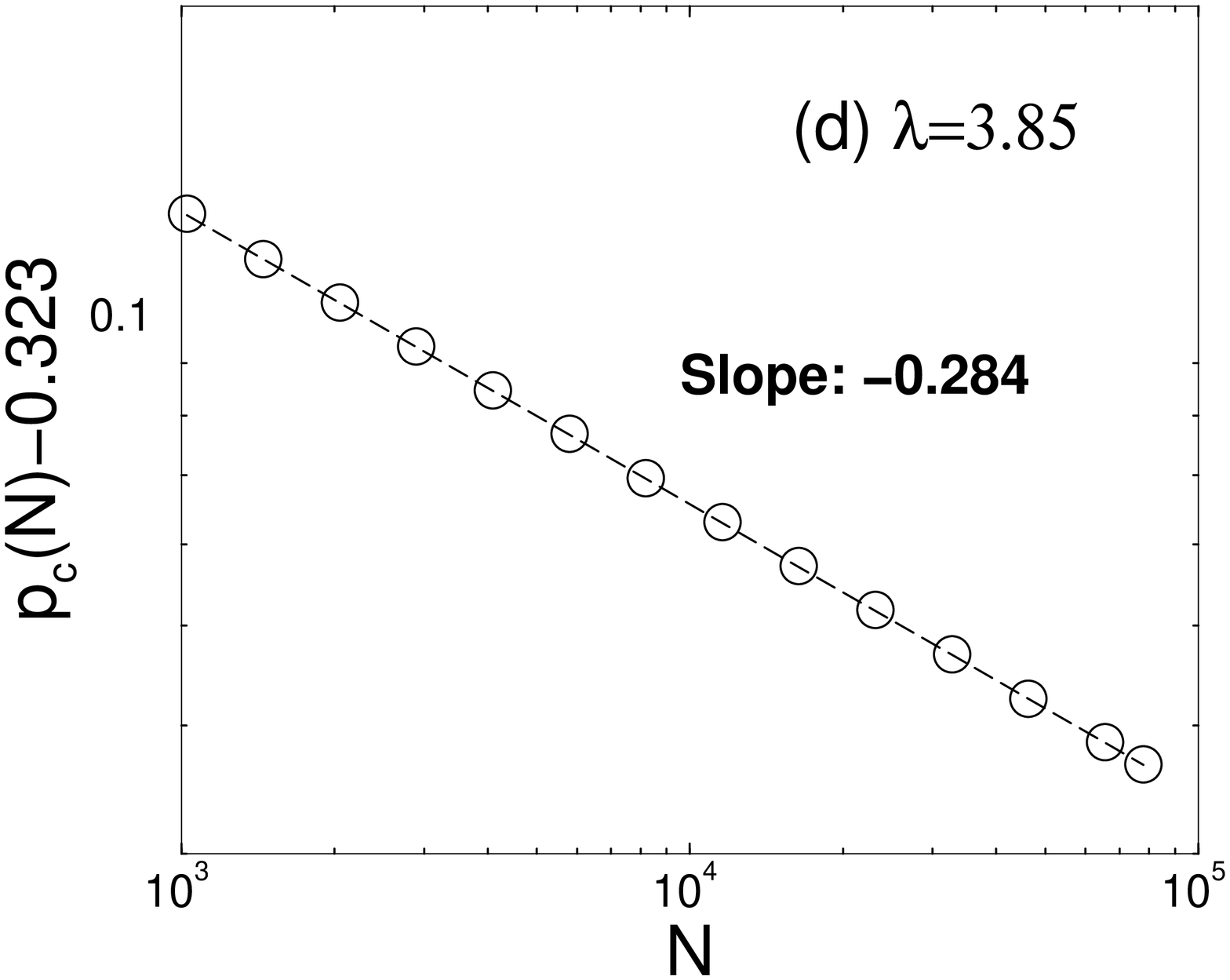}
  \includegraphics[width=0.19\textwidth, angle=-90]{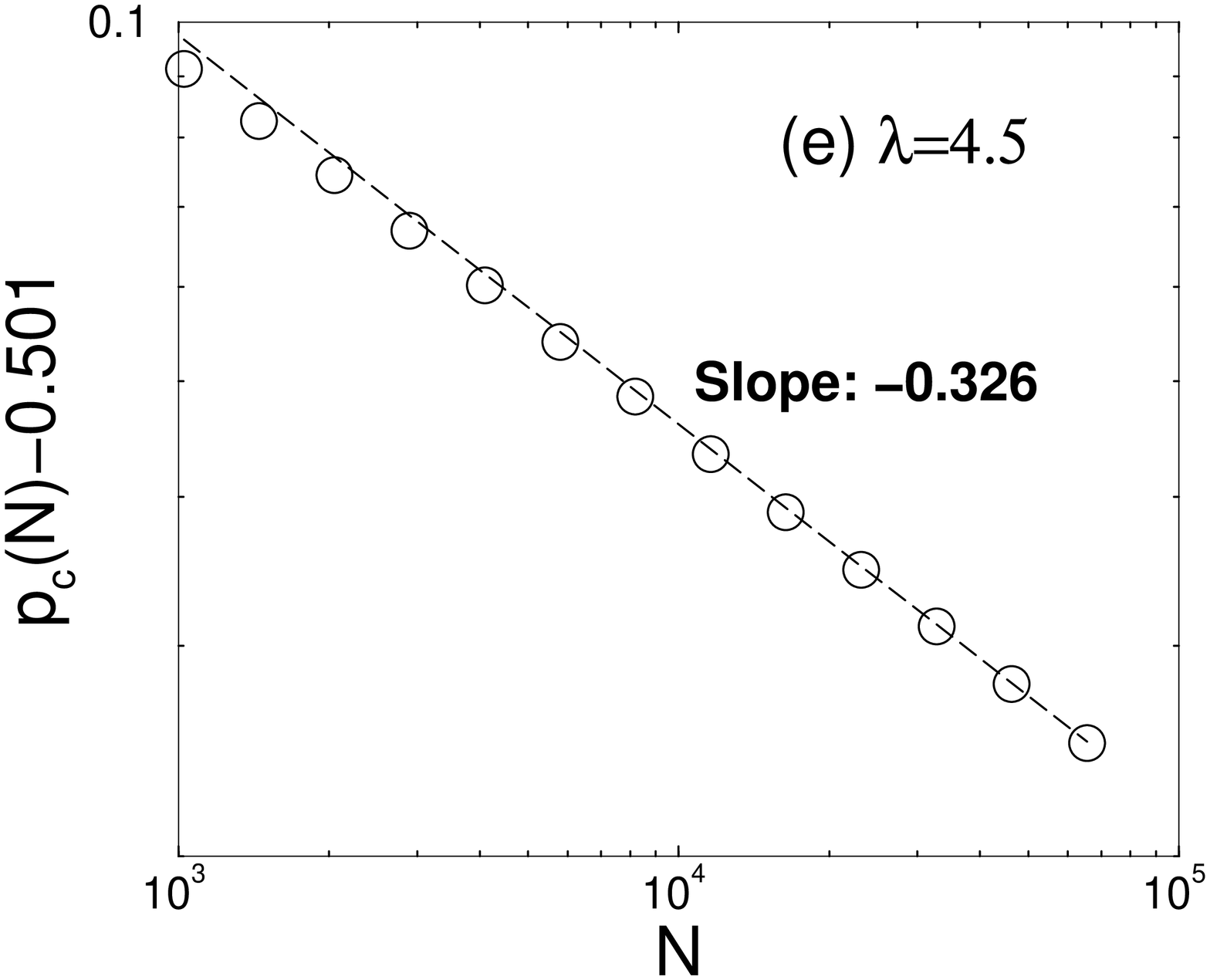}
  \caption{Log-log plots of $p_c(N) - p_c(\infty)$ as a function of $N$ for
    SF networks with $k_{\rm min} = 2$ and different value of $\lambda$. The
    dashed line is the reference line with indicated slope.}
  \label{pcN_sf.graph}
\end{figure}

\begin{figure}
  \includegraphics[width=0.19\textwidth, angle=-90]{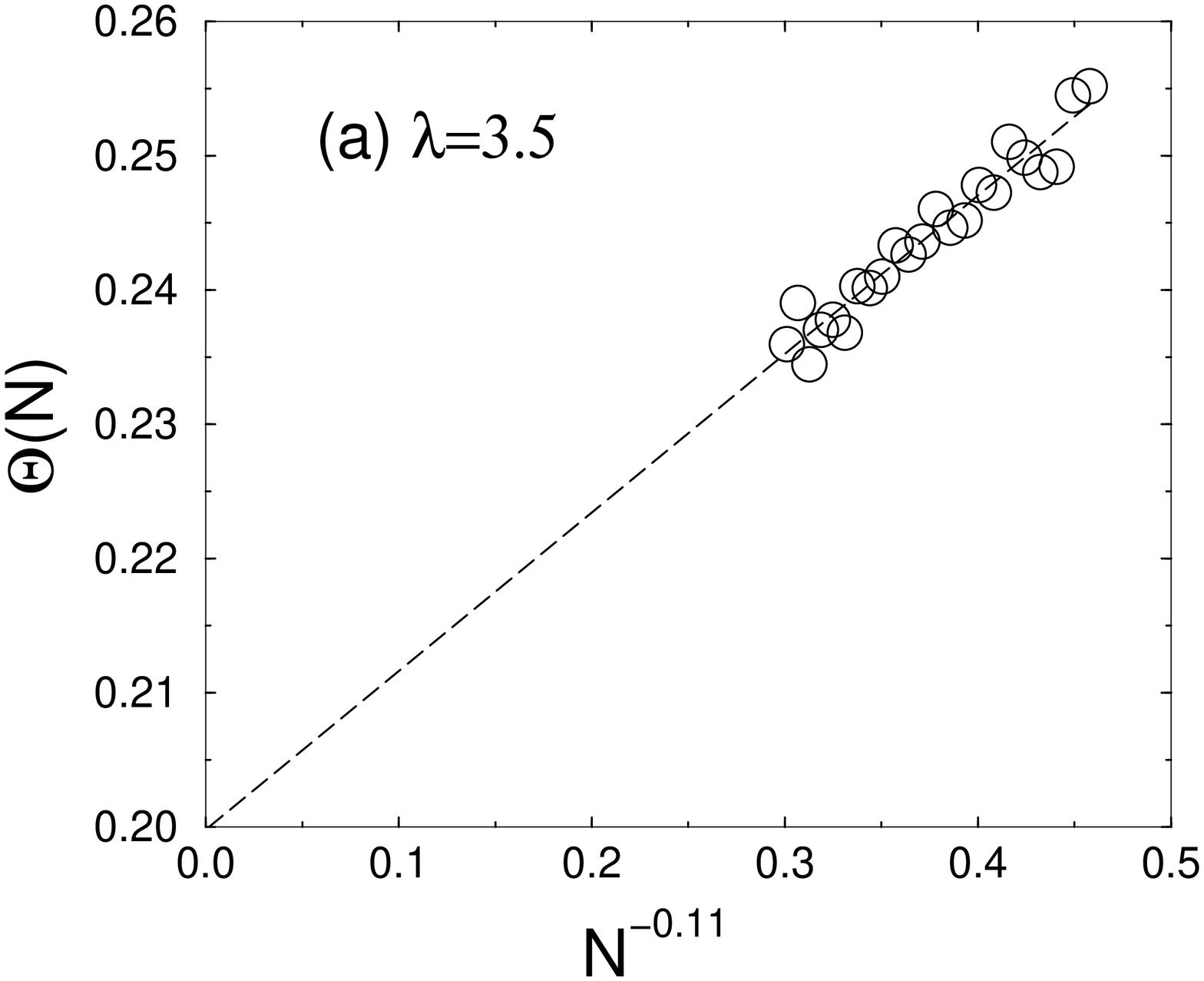}
  \includegraphics[width=0.19\textwidth, angle=-90]{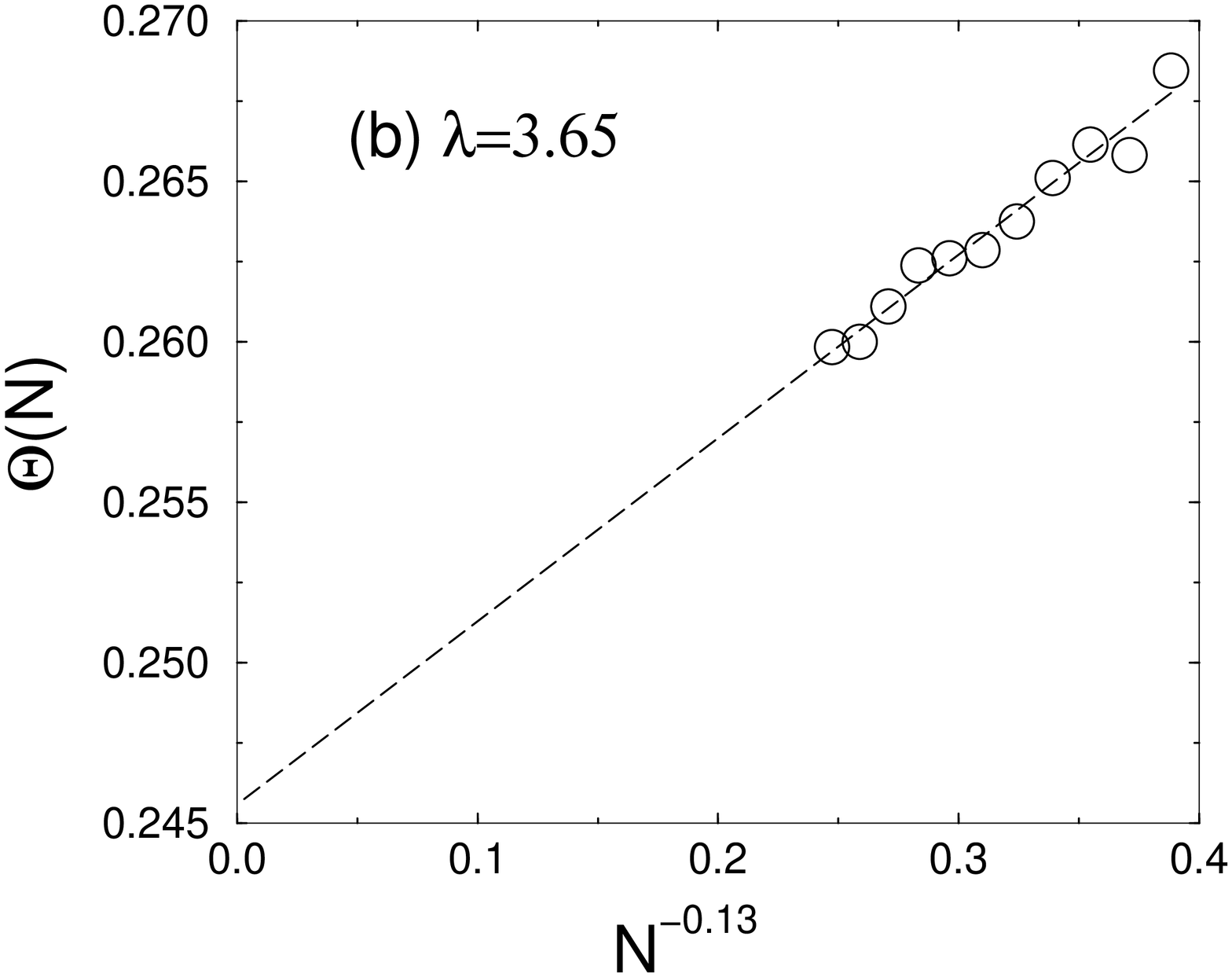}
  \caption{The exponent $\Theta(N)$ as a function of $N^{-x}$ for SF networks
  with $k_{\rm min} = 2$ and different value of $\lambda$: (a) $\lambda =
  3.5$, where $x \approx 0.11$; and (b) $\lambda = 3.65$, where $x \approx
  0.13$. The theoretical values $\Theta(\infty) = 0.2$ ($\lambda = 3.5$) and
  $\Theta(\infty) = 0.245$ ($\lambda = 3.65$), are consistent with the
  asymptotic values of $\Theta$ obtained for $N \to \infty$.}
  \label{theta_sf.graph}
\end{figure}

We calculate $\langle S_2 \rangle$ for SF networks for different values of
$\lambda$ and $N$ and compute $p_c(N)$ by fitting with a Gaussian function
near the peak of $\langle S_2 \rangle$ as for ER networks. Using the values
of $p_c(\infty)$ for SF networks displayed in Table~\ref{table_results}, we
obtain $\Theta$ by a power law fitting with Eq.~(\ref{finite_size_scale}) as
shown in Fig.~\ref{pcN_sf.graph}. As we can see for $\lambda = 4.5, 3.85$ and
$3.75$ we obtain quite good agreement with the theoretical values. However
for $\lambda = 3.65$ and $3.5$, the values of $\Theta$ become better when
fitting only the last several points (largest $N$) and still have large
deviations from their theoretical values. This strong finite size effect is
probably since for $\lambda \to 3$ the largest percolation cluster at the
criticality becomes smaller~\cite{FN_smax}. Thus, we expect that as $N$
increase, the exponent $\Theta(N)$ obtained by simulations should approach
the theoretical value of $\Theta$ of Eq.~(\ref{finite_size_scale_detail}). To
better estimate $\Theta$ we assume finite size corrections to scaling for
Eq.~(\ref{finite_size_scale_detail}), i.e.,
\begin{equation}
  p_c(N) - p_c(\infty) \sim N^{-\Theta}( 1 + N^{-x} ).
\end{equation}
Thus, the actual $\Theta(N)$ obtained from simulation is the succesive
slopes,
\begin{equation}
  \Theta(N) \equiv - \partial{\ln (p_c(N) - p_c(\infty) )}/\partial(\ln N),
\end{equation}
from which we can see that $\Theta(N)$ approaches $\Theta$ as a power law,
\begin{equation}
  \Theta(N) - \Theta \sim N^{-x}.
\end{equation}
Indeed, Fig.~\ref{theta_sf.graph} shows the exponent $\Theta(N)$ as a
function of $N^{-x}$ for $\lambda = 3.5$ and
$3.65$. Figure~\ref{theta_sf.graph}(a) shows that for $\lambda = 3.5$ and
$x=0.11$, we obtain a straight line and $\Theta(N)$ approaches $0.2$ as $N
\to \infty$, consistent with the theoretical value of $\Theta$
(Table~\ref{table_results}). Fig.~\ref{theta_sf.graph}(b) shows, for $\lambda
= 3.65$ and $x=0.13$, $\Theta(N)$ is again a straight line that approaches
$0.245$ for $N \to \infty$, consistent with the theory.

\begin{figure}
  \includegraphics[width=0.23\textwidth, angle=-90]{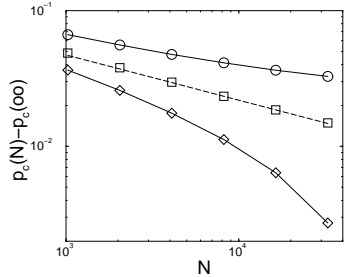}
  \caption{Log-log plot of $p_c(N) - p_c(\infty)$ as a function of $N$ for SF
  networks with $\lambda = 2.5$, $k_{\rm min} = 2$ for a targeted attack. The
  dashed line is the best fit with slope $-0.33$. Since we do not have a good
  estimation for $p_c(\infty)$, we modified $p_c(\infty)$ to get the best
  straight line in log-log plot, $p_c(\infty) = 0.23$ ($\circ$), $p_c(\infty)
  = 0.25$ ($\Box$) and $p_c(\infty) = 0.26$ ($\Diamond$). When $p_c(N) -
  p_c(\infty)$ is linear (dashed line) in the log-log plot, the slope yields
  the exponent $\Theta \approx 0.33$ i.e., $d_c = 6$.}
  \label{hub_kill_dc.graph}
\end{figure}

Next we estimate the value of $d_c$ for SF network under targeted attack on
the largest degree nodes~\cite{Callaway_robustness, Cohen_attack,
wu_pre}. For this case since the hubs are removed we expect that for all
$\lambda > 2$, $d_c$ will be the same as for ER, i.e., $d_c = 6$. In
Fig.~\ref{hub_kill_dc.graph}, we plot $p_c(N) - p_c(\infty)$ for SF with
$\lambda = 2.5$ under targeted attack. Indeed from
Eq.~(\ref{finite_size_scale}) by changing $p_c(\infty)$ and fitting the best
straight line in log-log plot, we obtain $\Theta \approx 0.33$, i.e., $d_c
\approx 6$, as expected.

\begin{figure}
  \includegraphics[width=0.25\textwidth, angle=-90]{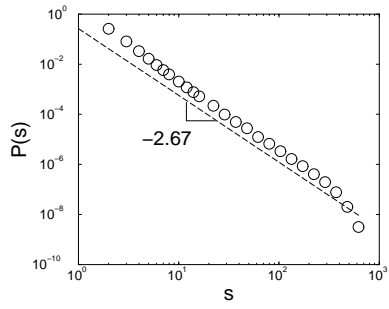}
  \caption{The probability distribution of the cluster sizes at $p_c(N)$ for
    $N=2048$ ($\circ$) and $N=16384$ ($\Box$). The dashed line is the
    reference line with slope $-2.67$.}
  \label{tau.graph}
\end{figure}

Further supports of the analytical aproach, we evaluat by simulations ${\cal
P}(s)$, the probability distribution of the cluster sizes at $p_c(N)$, which
should follow a power law for SF networks~\cite{Cohen_exponents},
\begin{equation}
  {\cal P}(s) \sim s^{-\tau} = s^{- (2 + \frac{1}{\lambda -2})}, 2 < \lambda < 4.
    \label{tau_def}
\end{equation}
Figure~\ref{tau.graph} shows the simulations results for SF networks $\lambda
= 3.5$. The dashed line is the reference line with slope $-2.67$, which is
the theoretical value of $\tau$ from Eq.~(\ref{tau.graph}), showing good
agreement between theory and simulations.

We thank ONR, ONR-Global, UNMdP, NEST Project No. DYSONET012911 and the
Israel Science Foundation for support. L. A. B. thanks to FONCyT (PICT-O
2004/370) for financial support. C. L. thanks to Conicet for financial
support.


\begin{table}[!h]
  \begin{tabular}{|c|c|c|c|}
    \hline $\lambda$	& $p_c(\infty)$	& Theoretical $\Theta$	& Numerical $\Theta$ \\
    \hline 3.50		& 0.2039	& 0.200			& 0.234  \\
    \hline 3.65		& 0.2574	& 0.245			& 0.260 \\
    \hline 3.75		& 0.2911	& 0.273			& 0.275 \\
    \hline 3.85		& 0.3234	& 0.298			& 0.284 \\
    \hline 4.50		& 0.5009	& 1/3			& 0.326 \\
    \hline ER ($\langle k \rangle = 4$)		& 0.25		& 1/3			& 0.328 \\
    \hline
  \end{tabular}
  \caption{The main results for SF and ER networks. The critical percolation
    threshold $p_c(\infty)$ indicates the numerical value calculated
    according to Eqs.~(\ref{pc_sf_cohen}) and
    (\ref{pk_modified}). Theoretical $\Theta$ is the theoretical prediction
    of $\Theta$ (from Eqs.~(\ref{ucd_sf})) and (\ref{finite_size_scale}) and
    numerical $\Theta$ is the numerical value we obtained from
    simulations. The SF networks were generated with $k_{\rm min} = 2$.}
  \label{table_results}
\end{table}


\newpage

\begin{thebibliography}{50}
\bibitem{FN_1} The case of $2 < \lambda < 3$ is more complex since there is
  no finite percolation threshold ($p_c \to 0$). However, since for $\lambda
  \to 3$, $d_c \to \infty$, Eq.~(\ref{ucd_sf}) suggests that for $2 < \lambda
  \le 3$ there is no finite upper critical dimension but $d_c = \infty$ for
  all $\lambda \leq 3$.
\bibitem{FN_kappa} The finite size dependency of
  Eq.~(\ref{finite_size_scale}) cannot be obtained from the condition $\kappa
  \equiv \langle k_0^2 \rangle / \langle k_0 \rangle = 2$, since this formula
  is valid only in the limit $N \to \infty$.
\bibitem{FN_smax} The largest percolation cluster at the criticality was
  analytically predicted as $\langle S \rangle \sim N^{(\lambda - 3) /
  (\lambda -1)}$~\cite{Cohen_exponents}.
\end{thebibliography}
\end{document}